\documentclass[nohyper]{JHEP3}

\usepackage[dvips]{graphicx,epsfig}
\usepackage{amsmath}
\usepackage{amsfonts}   
\usepackage{amssymb}  
\usepackage{mathrsfs}
\usepackage{exscale}

\def\be{\begin{equation}}
\def\ee{\end{equation}}
\def\bea{\begin{eqnarray}}
\def\eea{\end{eqnarray}}

\title{Isocurvature modes and Baryon Acoustic Oscillations}

\author{Anna Mangilli \\ ICC-UB (Instituto de Ciencias del Cosmos at Universitad de Barcelona), Marti i Franques 1, Barcelona, Spain\\
\& Institute of Space Sciences (IEEC-CSIC), Fac. Ciencies, Campus UAB, Bellaterra   \\ \email{anna.mangilli@icc.ub.edu}}
\author{ Licia Verde \\ ICC-UB (Instituto de Ciencias del Cosmos at Universitad de Barcelona), Marti i Franques 1, Barcelona, Spain\\
\& Institute of Space Sciences (IEEC-CSIC), Fac. Ciencies, Campus UAB, Bellaterra\\ \&ICREA (Institucio' Catalana de Recerca i Estudis Avancats) \\ \email{licia.verde@icc.ub.edu}}
\author{Maria Beltran \\ Department of Astronomy and Astrophysics, The University of Chicago, 5640 S Ellis, Chicago IL 60637, USA \\ \email{beltran@uchicago.edu}}

\abstract{
The measurement of Baryonic Acoustic Oscillations from galaxy surveys is  well known to be a robust and powerful tool to constrain dark energy.
This method relies on the knowledge of the size  of the acoustic horizon at radiation drag  derived from Cosmic Microwave Background Anisotropy  measurements.
In this paper we quantify the effect of non-standard initial conditions in the form of an isocurvature component  on the determination of dark energy parameters from future BAO surveys. 
In particular, if  there is an isocurvature  component (at a level still allowed by present data)  but it is ignored in the CMB analysis, the  sound horizon and cosmological parameters  determination is biased, and,  as a consequence, future surveys may incorrectly suggest  deviations from a cosmological constant.
In order to recover an unbiased determination  of  the sound horizon and dark energy parameters, a component of isocurvature perturbations must be included in the model  when analyzing CMB data.  Fortunately, doing so does not increase parameter errors significantly.
}

\begin{document}
\section{Introduction}

   Born on the basis of simplicity, the adiabatic framework describing the initial conditions for the perturbations, consolidated its popularity when it was shown to be predicted by the simplest one-field inflationary model \cite{mukhanov-adiabatic,Brandenberger92}.
 Although this simplest and basic adiabatic picture is widely accepted and  provides an excellent fit to current data (e.g., \cite{KomatsuWMAP7}),  there is not a priori reason to discard different and more general initial conditions involving entropy isocurvature perturbations. Such models  also relay on motivated physical assumptions, e.g. models of inflation with more than one field \cite{Linde1985,Kofman1986, Mollerach1990,
Polarski1994, Langlois99, Peebles1999, Bartolo2001}, neutrinos isocurvature perturbations \cite{Bucher-general-2001}, axion dark matter 
\cite{Axenides:1983hj, Lindeaxion, SeckelTurner, hybrid, TurnerWilczek, LindeLyth, Lyth:1991, Shellard:1997, Kawasaki1995axion} or the curvaton scenario 
\cite{curvaton1, curvaton2, curvaton3}.

 Even if pure isocurvature models have been ruled out \cite{Efstathiou86, Enqvist02, Page03, Hinshaw06}, current observations allow for mixed adiabatic and isocurvature contributions \cite{KomatsuWMAP7, Beltran-04, Beltran-05,Valiviita09,Dunkley05,Keskitalo07}. 
As shown in \cite{trotta2003, Valiviita09, Kurki05,Langlois00, Valiviita03, Bucher00,Sollom09}, the initial conditions issue is a very delicate problem: in fact, for current cosmological data, relaxing the assumption of adiabaticity reduces our ability to do precision cosmology since it compromises the accuracy of parameter constraints. 
 Generally, allowing for isocurvature modes introduces new degenerations in the parameters space which weaken considerably constraints.
  
 The Cosmic Microwave Background radiation (CMB), being our window on the early universe, is the preferred 
 data set to learn about initial conditions. 
 Up to now, however, the CMB temperature power spectrum alone, which is the CMB observable better constrained so far, has not been able to  break the degeneracy between the nature of initial  perturbations (i.e. the amount  and properties of an  isocurvature component) and cosmological parameters,
 e.g. \cite{ Kurki05, Valiviita03, Bean2006,Trotta01}.
 Even if the precision measurement of the CMB first acoustic peak at $\ell \simeq 220$ \cite{Page03,Hinshaw06} ruled out the possibility of a dominant isocurvature mode, allowing for isocurvature perturbations together with the adiabtic ones introduce additional degeneracies in the interpretation of the CMB data that current experiments could not break.
 Adding external data sets somewhat alleviates the issue for some degeneracy directions e.g. \cite{Beltran-04,Dunkley05}.
  As shown in \cite{bucher-pol}, the precision polarization measurement of the next CMB experiments like Planck will be crucial to lift such degeneracies, i.e., to distinguish the effect of the isocurvature modes from those due to the variations of the cosmological parameters.

 It is important to keep in mind that analyzing the CMB data with the prior assumption of purely adiabatic initial conditions when the real universe contains  even a small  isocurvature contribution, could lead to an incorrect determination of the cosmological parameters.  In fact,  the presence of an isocurvature component changes the shape and the location of the CMB acoustic peaks, mimicking the effect of  parameters such as $\Omega_m h^2$, $H_0$ and $w$ (see \cite{Valiviita09} and references therein).
 
 Here we investigate wether such an effect has an impact on ``standard rulers'', like the sound horizon at recombination, inferred from  CMB observations. 
   
This issue turns out to be very delicate and important especially in view of 
 the next generation of galaxy surveys which aims at probing with high accuracy the late time expansion and thus the nature of dark energy by means of Baryon Acoustic Oscillations (BAO) at low redshift ($z<2$). 
 
 The BAO in the primordial photon-baryon fluid leave in fact an imprint in the large scale matter distribution and, at each redshift,  their physical properties are related to the size of the sound horizon at radiation drag (i.e., when baryons were released from the photons) e.g. \cite{zd-fit-eis-hu}. 
Thus it is possible to use measurements of these oscillations at different redshifts as standard rulers.
The key idea is that measuring the BAO signal at low redshift with large scale structure surveys and knowing the size of the sound horizon at radiation drag %
with high precision from the CMB should allow to probe %
the expansion history, i.e., the Hubble parameter $H(z)$,  and the angular diameter distance $D_A(z)$ at different redshift \cite{Seo-Eisenstein} and thus the dark energy properties.

It has been found that the effects of  a number of theoretical systematics  such as non-linearities, bias etc, on the determination of the BAO location can be minimized to the point that BAO are one of the key observables of the next generation dark energy experiments.  It is therefore important  to investigate the robustness of the method to other theoretical uncertainties.  A crucial assumption is the possibility to measure accurately and robustly the acoustic sound horizon at radiation drag.   Ref \cite{EisensteinWhite04} showed that generally CMB observations yield a robust determination  of such standard ruler although there could be possible systematic effects introduced by deviations from the standard evolution of the early universe. Refs \cite{deBernardis09, Linder08} have considered two early universe non-standard phenomena that  change  the standard recombination process and  could affect BAO interpretation. 
 In this paper we investigate  another possible deviation from  the minimal, standard cosmological model that  could affect the BAO interpretation as standard rulers if not taken into account:
 the presence of an isocurvature contribution to the initial conditions for primordial perturbations.  This point constitutes the new part of this work, since such an effect has not been explored or quantified in details so far.

  The paper is organized as follows: in the next section 
we give a brief introduction to isocurvature modes and the notation used. This is only review material but it is reported here to clarify our set-up and notation. In section \ref{sec:analysis} we investigate the effects on  forecasted parameter constraints for an experiment with characteristic similar to the Planck satellite, arising from the presence of an isocurvature contribution to the primordial perturbations.  (In what follows we will refer to such an experiment as ``Planck'' for  short). We quantify  the systematic effects  on quantities like the sound horizon at radiation drag and the expansion history parameter, introduced if an isocurvature component at a level  allowed by current data,  was present in the data but ignored in the analysis.

  In section \ref{sec:results} we present our results on the impact of isocurvature contributions on the BAO observables and we explore the implications on dark energy constraints from future surveys forecasts. Finally in section \ref{sec:conclusions} we present our conclusions. Appendix A reports the relevant equations and definitions for BAO, in Appendix B we recap the basic equations for the Fisher matrix approach to forecasting and Appendix C explores the source of CMB parameters degeneracies introduced by  an  extra parameter for the isocurvature  contribution to the  initial perturbations.
  

\section{Isocurvature: some theory and notation}

  The simplest characterization of  the primordial perturbations \footnote{In this context primordial perturbations refers to perturbations defined deep in the radiation era on superhorizon wavelengths with modes initially excited well before recombination so that any decay mode had time to vanish and the main cosmological components are: photons, baryons, neutrinos and CDM.} is described by the adiabatic framework: 
all particles species are perturbed in spatially uniform ratio so that the relative ratios in the number density remain unperturbed.
 Given two different particle species, X and Y, the adiabatic condition requires:
\be
\delta\left(\frac{ n_X}{n_Y}\right)=0,
\ee
where $n_X$ and $n_Y$ are the correspondent particle number densities.

The global perturbation on the matter component, because of the Einstein equations, is associated with a curvature perturbation. This is the reason why adiabatic perturbations are also called curvature perturbations.

However, this is not the only possibility. There can exist in fact perturbations associated with fluctuations in number density between different components of the cosmological plasma  in the  early universe, well before photon decoupling. These are called isocurvature or entropy perturbations and are generated by the stress fluctuations through the causal redistribution of matter under energy-momentum conservation. Causality and momentum conservation require that the initial curvature perturbations must vanish (spatially varying abundance of particles species are arranged to cancel locally leaving the curvature of the spatial hypersurface unperturbed).
Density perturbations are then produced by non-adiabatic pressure (entropy) perturbations. See \cite{Hu-Sper-Whi97} for a detailed description. 

The entropy perturbation between two particle species (i.e., fluid components) can be written in terms of the density contrast $\delta_i=\frac{\delta \rho_i}{\rho_i}$ and the equation of state parameter $w_i$ as \footnote{Recall that the continuity equation, which follows from the energy-momentum conservation, takes the form $\dot{\rho} = -3 H (P + \rho)=- 3H\rho (1+w)$.}:
\be
{\cal S}_{XY}= \frac{ \delta_X}{1+ w_X} - \frac{ \delta_Y}{1 + w_Y}.
\ee
 Formally, this is equivalent to: ${\cal S}_{XY}= \frac{ \delta n_X}{n_X} - \frac{ \delta n_Y}{n_Y}$, 
which quantifies the variation in the particle number densities between two different species.

More precisely (see \cite{Bucher-general-2001} and references therein) the most general description of a primordial perturbation accounts for 5 non-decaying (regular) modes corresponding to each wavenumber: an adiabatic (AD) growing mode, a baryon isocurvature mode (BI), a cold dark matter isocurvature mode (CDI), a neutrino density (NID) and a neutrino velocity (NIV) mode.
For the dark matter component $\frac{\delta n_c}{n_c}=\frac{\delta \rho_c}{\rho_c}$, so that the CDM isocurvature mode can be written as:
\be
{\cal S}_c \equiv \delta_c - \frac{3}{4} \delta_\gamma,
\ee
where $\delta_X=\frac{\delta \rho_X}{\rho_X}$ is the energy density contrast of the X particle species.  Analogously the baryon (b) and the neutrinos ($\nu$) isocurvature modes take the form: 
\bea
{\cal S}_b \equiv \delta_b - \frac{3}{4} \delta_\gamma \\
{\cal S}_\nu \equiv \frac{3}{4}  \delta_\nu - \frac{3}{4} \delta_\gamma.
\eea
Of course for the adiabatic mode $S_c=S_\nu=S_b=0$.
	
One of the crucial distinctions between the adiabatic and the isocurvature models relies on the behavior of the fluctuations at very early time during horizon crossing at the epoch of inflation (or analogous model for the very early universe). In the adiabatic case constant density perturbations are present initially and imply a constant curvature on super-horizon scales, while in the pure isocurvature case there are not initial density fluctuations which instead are created from stresses in the radiation-matter component.

\subsection{A worked example: the curvaton model}

An alternative to the standard single field inflationary model is the curvaton scenario \cite{curvaton1,curvaton2,curvaton3}. 
This model relies on the inflaton, the light scalar field that dominates the background density during inflation, and the curvaton field $\sigma$ which, decaying after inflation, seeds the observable cosmological perturbations. 
Within this framework, the initial conditions then correspond to purely entropy primordial fluctuations,
namely isocurvature perturbations, because the curvaton field practically does not contribute to perturbations of the metric. 
If the field $\sigma$ decays sufficiently early and transforms completely into thermalized radiation, pure adiabatic perturbations with a blue spectrum are generated and the initial isocurvature density perturbations disappear. 

In general, by tuning the decay dynamics of $\sigma$, the curvaton scenario allows for mixed adiabatic and isocurvature fluctuations, 
with any residual isocurvature perturbation correlated or anti-correlated to the adiabatic density one and with the same tilt for the both spectra: $n_{ad}=n_{iso}$.

Besides the fact that  current data are compatible with this theoretical picture, the curvaton model attracted growing attention because it predicts primordial non-gaussianity features of the local type in the spectrum of primordial perturbations. In the paper we use the curvaton scenario as a working example for a model that  gives rise to a small fraction of  correlated CDI isocurvature. As shown in the next sections, our analysis can be generalized to models with an arbitrary amount and type of isocurvature. The curvaton has  $n_{ad}=n_{iso}$, therefore our findings are quantitative only for this case: we will comment on how our findings should  qualitatively hold also for $n_{ad}\ne n_{iso}$.

\subsection{Isocurvature and the cosmological parameter $f_{iso}$}

A common parametrization for the isocurvature perturbations is given by \cite{Peiris03}: 
\be\label{e:fiso}
f_{iso}=\frac{\langle {\cal{S} }^2_{rad} \rangle^{1/2}}{\langle {\cal{R}}^2_{rad} \rangle^{1/2}},
\ee
defined as the ratio between the entropy $ {\cal{S}}$ and the curvature (adiabatic) $ {\cal{R}}$ perturbations evaluated during the radiation epoch %
and %
at a pivot scale $k_0$.  In our case  we set %
$0.05 \; Mpc^{-1}$, as often done in the literature. 
The correlation coefficient can be then defined in terms of an angle $\Delta_{k_0}$ such that:
\be
\cos \Delta_{k_0} =\frac{\langle  {\cal{R}}_{rad} {\cal{S} }_{rad} \rangle}{\langle {\cal{R}}^2_{rad} \rangle^{1/2} \langle {\cal{S}}^2_{rad} \rangle^{1/2}}.
\ee
Throughout this paper we will use a sign convention such that $f_{iso}<0$ will correspond to correlated modes.
As pointed out in \cite{Kurki05},  in a practical implementation of a multi-parametric analysis using Markov Chains Monte Carlo (MCMC), the $f_{iso}$ parametrization (instead of $f^2_{iso}$) favors small multiplier in front of the isourvature component thus adding an implicit bias towards pure adiabatic model in the posterior.
Another parametrization in terms of the amplitude parameter $\alpha$ for which the data have a linear response is given by:
$\alpha=\frac{f^2_{iso}}{(1+f^2_{iso})}, \; \beta=\cos(\Delta_{k_0})$,
where maximally correlated (anticorrelated) modes correspond to $\beta=+1$ ( $\beta=-1$).  Our analysis is mostly concerned with degeneracies directions and is thus less sensitive to the effect of the prior choice on the posteriors.

For the curvaton scenario with mixed adiabatic and purely correlated CDM isocurvature modes which we chose as working example, the most recent constraints are  $\alpha_{-1} < 0.011$ (95\% CL), \cite{KomatsuWMAP7}.

  \begin{figure*}
\centering
\includegraphics[width=13cm]{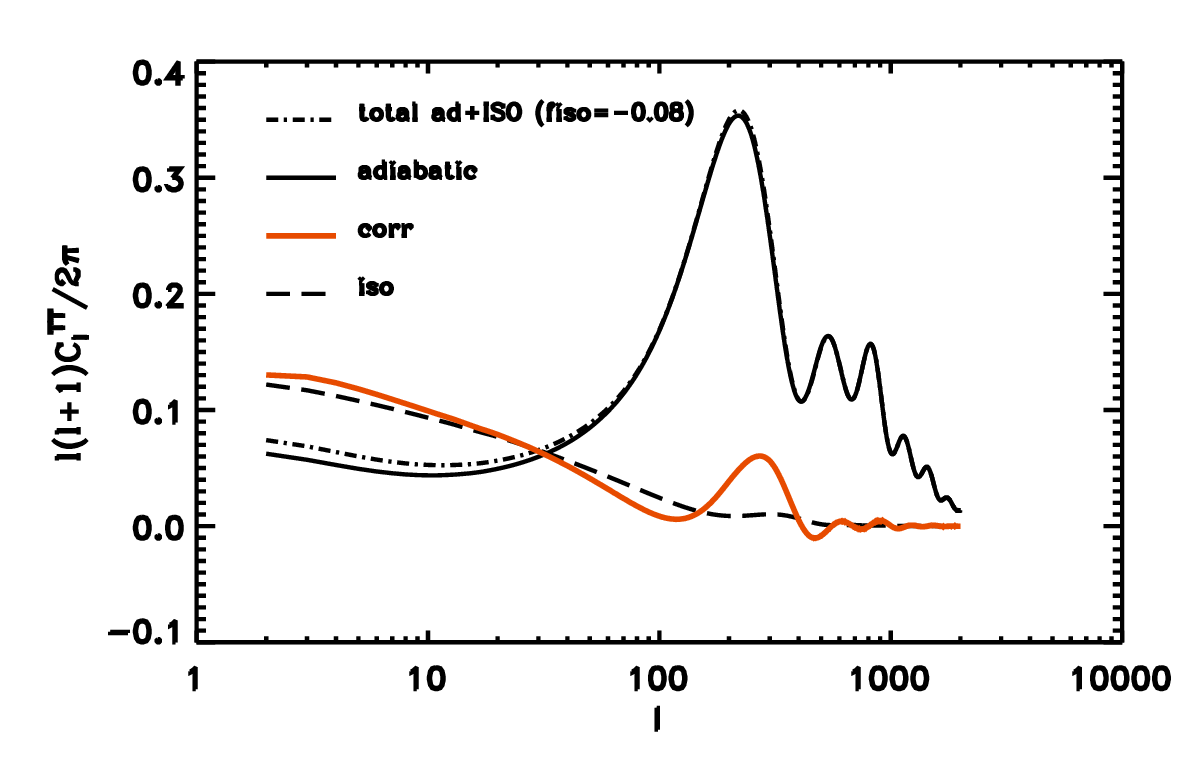} 
\caption{The different components of the unit-amplitude temperature angular power spectrum defined in Eq. \ref{eq:Cls-tot} (dot-dashed line) for the cosmological parameters listed in table \ref{t:fid} and in the case $n_{iso}=n_{ad}$: $C_\ell^{AD}$ (solid black line), $C_\ell^{ISO}$ (dashed line) and $C_\ell^{cor}$ (orange line). See Eqs. (\ref{eq:Cls-ad} - \ref{eq:Cls-corr}). 
}
\label{fig:Cls}
\end{figure*}

 Since we are interested in the effect of adding an isocurvature contribution on the CMB angular power spectrum it is useful to give the explicit expression for this. In general the two-point correlation function or power spectra for the adiabatic mode, the isocurvature mode and their cross-correlation can be described by %
  two amplitudes, one correlation angle and three independent spectral indexes ($n_{ad}$, $n_{iso}$ and $n_{cor}$), so that, in the case of the CMB, the respective power spectra are:
\be\label{eq:Cls-ad}
C_\ell^{ad}=\int \frac{dk}{k}[\Theta^{ad}_\ell(k)]^2 \left( \frac{k}{k_0} \right)^{n_{ad}-1},
\ee

\be\label{eq:Cls-iso}
C_\ell^{iso}=\int \frac{dk}{k}[\Theta^{iso}_\ell(k)]^2 \left( \frac{k}{k_0} \right)^{n_{iso}-1}
\ee
and
\be\label{eq:Cls-corr}
C_\ell^{cor}=\int \frac{dk}{k} \Theta^{ad}_\ell(k)  \Theta^{iso}_\ell(k) \left( \frac{k}{k_0} \right)^{n_{cor}+\frac{1}{2}( n_{ad}+n_{iso})-1}.
\ee
Here $\Theta^{ad}_\ell(k)$ and $ \Theta^{iso}_\ell(k)$ are the radiation transfer functions for adiabatic and isocurvature perturbations that describe how an initial perturbation evolved to a temperature or polarization anisotropy multipole $\ell$.
The total angular power spectrum takes then the form:
\be\label{eq:Cls-tot}
C_\ell=\langle {\cal{R} }^2_{rad} \rangle [C_\ell^{ad}+f^2_{iso} C_\ell^{iso} + 2 f_{iso} \cos \Delta_{k_0} C_\ell^{cor}].
\ee
Recall that in this paper we will always work in the case where $n_{ad}=n_{iso}$. In Fig.  \ref{fig:Cls} we show the unit-amplitde components of the temperature angular power spectrum listed in the above equations: $C_\ell^{AD}$ (solid black line), $C_\ell^{ISO}$ (dashed line) and $C_\ell^{cor}$ (orange line). The cosmological parameters used are given in tab. \ref{t:fid} and in the total mixed ad+iso power spectrum (dot-dashed line) the isocurvature fraction is $f_{iso}=-0.08$. 
 From the plot and from eq. (\ref{eq:Cls-tot}) %
 it is clear that, for small isocurvature fractions, the main isocurvature contribution come from the mixing term %
 coefficient $2 f_{iso} \cos \Delta_{k_0}$.

\section{Analysis}\label{sec:analysis}

The aim of this section is to investigate and quantify the following.
Assume that the Universe contained a small  isocurvature  contribution  to the primordial perturbation at a level allowed by current constraints. If forthcoming Planck satellite data were to be analyzed in the context of purely adiabatic initial conditions: {\it a)} what would be the systematic error on quantities like the sound horizon at radiation drag and what would be the implications for  the  reconstructed expansion history from BAO observations? (sec. \ref{s:mcmc})
{\it b)} what would be the shift on cosmological parameters and on the sound horizon at radiation drag as a function of the amount of   isocurvature  contribution? How does this shift  compare with the forecasted statistical errors on the same quantities? (sec. \ref{sec:Fisher}).

Throughout we will  assume a flat universe. Recall however, as pointed out in \cite{Valiviita09, Dunkley05}, that assuming spatial flatness in isocurvature studies strongly biases the result toward pure adiabaticity.

\subsection{MCMC  approach to forecasts  for a Planck-like experiment}\label{s:mcmc}

We adopt  a modified version of the CAMB code \cite{CAMB,Beltran-04,Beltran-05} which includes the initial conditions for the correlated mixed adiabatic (AD) and cold dark matter isocurvature (CDI) modes predicted by the curvaton scenario.  With this code we generate two fiducial models: an adiabatic one and one with an  isocurvature   contribution still allowed by current data, as summarized in  Table \ref{t:fid}:

\begin{itemize}
\item Fid.AD-  Fiducial model Adiabatic (AD) - 7 parameters.
\item Fid.ISO- Fiducial model curvaton - 8 parameters.
\end{itemize}
The set of parameters we consider is given by: $\omega_b=\Omega_b h^2$, $\omega_c=\Omega_c h^2$ that are respectively the baryon and cold dark matter physical density fractions, the optical depth at reionization $\tau$, the Hubble parameter $H_0=100 \, h$Km sec $Mpc^{-1} $, the scalar amplitude $A_s=ln[10^{10}{\cal R}_{rad}]$ , being ${\cal R}_{rad}$ the curvature perturbation during the radiation era,  the scalar adiabatic tilt $n_s$, the dark energy equation of state parameter  $w$ and $f_{iso}$ the isocurvature fraction.

We estimate Planck satellite errors on the CMB temperature and polarization and the form for the likelihood using the specifications of Ref. \cite{bluebook, Baumannetal} and  the technique presented in \cite{verdepol05}.

We use the  code cosmoMC  \cite{cosmoMC}  to sample the posterior distribution for the cosmological parameters and a number of derived parameters using the Markov Chain Monte Carlo (MCMC) method.

\begin{table*}%
\caption{Adiabatic and isocurvature fiducial models parameters.
	}
\begin{center}
\begin{tabular}{|ccc|} 
\hline
\hline
Parameter&Ad-$\Lambda$CDM&Isocurvature\\
\hline
$\omega_b$&0.022&0.022\\
$\omega_c$&0.11&0.11\\
$h$&0.704&0.704\\
$\tau$&0.073&0.073\\
$n_s$&0.96&0.96\\
$A_s$&$2.4á10^{-9}$&$2.4á10^{-9}$\\
$w$&$-1$&$-1$\\
$f_{iso}$&0&$-0.08$\\
\hline
\end{tabular}\label{t:fid}
\end{center}
\end{table*}

Each fiducial  is sampled with two different models:

\begin{itemize}
\item fit.AD Model with parameters: $\{\omega_b,\omega_c,\theta, \tau, n_s,A_s,(w)\} \, (f_{iso}=0)$
\item fit.ISO Model with params: $\{\omega_b,\omega_c,\theta, \tau, n_s,A_s,(w),f_{iso} \}$.
\end{itemize}
A flat prior on these parameters is assumed.
As usual  the variable $\theta=100 \frac{r_s(z*)}{D_A(z*)}$ is used  (and a flat prior on this variable is assumed) instead of $H_0$ or $\Omega_\Lambda$ since it is a ``physical'' parameter and improves MCMC performance; $r_s(z*)$ and $D_A(z*)$ are, respectively, the sound horizon and the angular diameter distance at decoupling (see Eqs \ref{eq:rsz} and \ref{eq:ang-dist}). Both the AD and ISO analyses are run once fixing $w=-1$ and once letting  $w \neq -1$ (but still keeping it constant in time). 

For our purpose we will mainly use the chains: {\it i}) fiducial model curvaton fitted with adiabatic (Fid.ISO-fit.AD) for both cases  cosmological constant ($w=-1$) and   $w\ne -1$  but constant (Fid.ISO-fit.AD$w$)  and  {\it b)} fiducial model adiabatic fitted with an anti-correlated isocurvature model (Fid.AD-fit.ISO) in the $w=-1$ case.  (The other cases Fid.ISO-fit.ISO and Fid.AD-fit.AD are used only for cross checks and for testing that our procedure is not biased). The  recovered best fit values  (N-dimensional  maximum likelihood location) for the cosmological parameters are listed in table \ref{t:getdist}. Only in the case where the mean  best fit model is different from the maximum likelihood one both are reported.

 The "goodness of fit'' parameter $Q$\footnote{
$Q(\nu, \hat \chi^2)=1- \Gamma(\nu/2,\hat \chi^2)$, where $\nu=n-m$ are the degrees of freedom (the difference between the number of independent data points $n$ and the number of parameters $m$) and $\hat \chi^2$ is the minimum $\chi^2$.
}
is $0.193871$ and $0.464991$, respectively for Fid.ISO-fit.AD and  Fid.AD-fit.ISO  assuming that Planck has 4000 independent data points (assumed to be the $C_{\ell}$ for temperature and polarization) so that both fits are acceptable. 

To begin with, we are 
interested in disentangling the effects of an (ignored) isocurvature contribution on an `early' type observable like the sound horizon at the 
radiation drag and on a `late' type observable like the expansion history at low redshift  ($0<z<2$). The acoustic oscillations imprinted in the galaxy distribution can in principle be measured at different redshifts so it is important to distinguish the effects of an isocurvature contribution on expansion-history observables  (e.g., $H(z)$) or determining the initial layout of the acoustic feature ($r_s(z_d)$).

\begin{table*}
\begin{center}
\begin{tabular}{|ccccc|} 
\hline
\hline
$\;\;\;$&Fid.ISO-fit.AD&Fid.ISO-fit.AD$w$&Fid.ISO-fit.AD$w$&Fid.AD-fit.ISO $^{a}$\\
&&best fit&mean&\\
\hline
$\Delta lnL$&68.997&67.784&-&0.198\\
$\omega_b$&0.02189&0.0219&0.02189&0.022\\
$\omega_c$&0.10549&0.105633&0.10551&0.110087\\
$n_s$&0.9784&0.97804&0.9783&0.9597\\
w&-1&-1.342&-1.171&-1\\
$\Omega_m$&0.249&0.171472&0.2113&0.26691\\
$\Omega_\Lambda$&0.757084&0.828528&0.7886&0.733089\\
$H_0$&72.4005&86.2415&79.52&70.3461\\
\hline
\end{tabular}\label{t:getdist}
\caption{MCMC-recovered maximum likelihood  parameter values (i.e. likestats file). For the varying $w$ chains we report the mean (.margestats file) parameters since in this case they differ from the maximum likelihood ones (in the other cases there is no noticeable difference). {\it a}: In this case the best fit model has a tiny anti-correlated isocurvature  model; the best fit gives a slightly negative $f_{iso}$ but well  within $1-\sigma$ error. }
\end{center}
\end{table*}

\subsubsection{Sound horizon at radiation drag $r_s(z_d)$}\label{rszd}

The sound horizon at radiation drag $r_s(z_d)$, i.e., the distance sound can travel up to the time when the baryons were released from the photons, is defined by Eq. (\ref{eq:rsz}). %
The radiation drag redshift  $z_d$, %
according to the fitting formula from \cite{zd-fit-eis-hu} used in the literature so far, can be explicitly written as:
\bea
z_d=\frac{1291 (\Omega_m h^2)^{0.251}}{1+0.659(\Omega_m h^2)^{0.828}}[1+b_1(\Omega_b h^2)^{b_2}]\\
b_1= 0.313 (\Omega_m h^2)^{-0.419} [1 + 0.607 (\Omega_m h^2)^{0.674}]\nonumber\\
b_2=  0.238(\Omega_m h^2)^{0.223}, \nonumber \;\;\;\;\;\;\;\;\;\;\;\;\;\;\;\;
\eea
with $\Omega_m=\Omega_b+\Omega_{cdm}$ the matter density parameter.

Actually this is an approximation and the drag epoch $z_d$ can be calculated more precisely, as pointed out in \cite{zd-camb-Hamann},  %
as the epoch at which the drag optical depth $\tau_d$ equals one: %
\be
\tau_d(z_d) \equiv \frac{3}{4}\frac{\omega_\gamma}{\omega_b} \int_0^{z_d} dz \frac{d \eta}{da} \frac{x_e(z) \sigma_T}{1+z} =1,
\ee
where $x_e(z)$ is the fraction of free electrons and $\sigma_T$ the Thomson cross-section.

\begin{table*}
\begin{center}
\begin{tabular}{|ccccc|} 
\hline
\hline
Type of chain&$z^{FIT}_d$&$z^{CAMB}_d$&$r_s(z^{FIT}_d)$&$r_s(z^{CAMB}_d)$\\
\hline
 Fid. Adiabatic&$1018.7948$&$1058.5975$&$154.198$&$150.449$\\
Fid.AD-fit.ISO&1018.9217&1058.559&153.088&150.430\\
Fid. Curvaton&$1018.8711$&$1058.52127$&$154.198$&150.449\\
Fid.ISO-fit.AD&$1018.1467$&$1057.910919$&$155.731$&151.888\\
Fid.ISO-fit.AD$w$ best fit&$1018.7948$&$1057.94906$&$153.803$&151.823\\
\hline
\end{tabular}\label{t:rszd}
\caption{Radiation drag redshift  $z_d$ and the sound horizon $r_s(z_d)$: 
we calculate the radiation drag redshift $z_d$ by using the fitting formula from \cite{zd-fit-eis-hu} and by using a modification of the CAMB code \cite{zd-camb-Hamann}: $z^{FIT}_d$ and $z^{CAMB}_d$. In the last two column we report the
sound horizon at radiation drag $r_s(z_d)$ (Mpc) for both cases. Note that (see sec. \ref{sec:Fisher}) the $1-\sigma$ error forecast for Planck on $r_s(z_d)$ is $0.359$ (Mpc). 
	}
\end{center}
\end{table*}

By using the best fit values for the parameters %
for each type of chain (table \ref{t:getdist}), we calculate the sound horizon at radiation drag numerically in both cases. For the second more precise calculation we used a modified version of CAMB to extract both $z_d$ and $r_s(z_d)$ directly from the code. We find that, the use of the  fitting formula yield  values for $r_s(z_d)$  inaccurate by up to  4\%.
The results are summarized in the table \ref{t:rszd}.
It is clear from the 7th column of table  \ref{t:rszd} that 
if the underlying fiducial model is the curvaton and it is fitted with an adiabatic model the systematic shift  in the estimated  value of $r_s(z_d)$ is $\simeq 1\%$ for both the $w=-1$ and  $w$ cases, while if the adiabatic fiducial model is fitted with an (anticorrelated) isocurvature model --as expected-- there is no systematic shift.  

Adding an isocurvature contribution implies a small variation in the estimation of the sound horizon at radiation drag that however should be taken into account for the Planck experiment since it is expected to measure such observable with high accuracy (better that 1\%) , as we will discuss in \S \ref{sec:Fisher}. The implication of this systematic shift on the interpretation for future BAO data is explored in \S \ref{sec:results}.

\subsubsection{Expansion history $H(z)$}

By using the best fit values for the parameters of each chain, 
 we calculate the expansion history $H(z)$
 up to redshift $z=2$, which is the redshift region of interest for studying the late-time expansion and accessible by current and future large-scale structure BAO surveys. Within this range and in the case of a flat model 
it takes the form:  
 \be
H(z)=H_0 \sqrt{\Omega_m (1+z)^3 + \Omega_\Lambda(1+z)^{3(1+w)}}.
\ee

We find that the variation in the  CMB-predicted expansion history $H(z)$ due to the fact that the underlying fiducial curvaton model is fitted with a pure adiabatic model is about 3\% for a $\Lambda$CDM model at $z=0$ while it can reach 25 \% for a model with varying dark energy equation of state $w$.

These results are a quantitative confirmation of what previously anticipated: working with the prior assumption of purely adiabatic initial conditions when the real universe contains an isocurvature contribution could lead to an incorrect determination of the cosmological parameters. In particular, the presence of an isocurvature fraction can affect the determination of the value of the sound horizon at the radiation drag. It also affects the CMB-recovered value  of the Hubble parameter at the present time $H_0$ and the expansion history --$H(z)$-- inferred from CMB analysis.

\subsection{Planck Fisher analysis}\label{sec:Fisher}
The MCMC approach is well suited to explore a single  case (for a particular choice of $f_{iso}$ value). To study the dependence of the effect illustrated in \S \ref{s:mcmc}  on the size of the isocurvature contribution, a Fisher matrix approach is  better suited.
The Fisher matrix analysis \cite{Fisher} (see \ref{app:fisher} for more details) set up for the Planck experiment is calculated for a fiducial model with fiducial  parameters:
\be\label{eq:fisher-params}
\{r, n_s, dn_s/d\ln k,z_{re}, \omega_b, \omega_c, h, A_s, f_{iso}\}=\{0.01,0.963, 0. ,0.84, 0.02273,0.1099,0.72, 0.8169, -0.01\},
\ee
where $n_s$, $dn_s/d\ln k$ and $z_{re}$ are, respectively, the spectral index, the running of the spectral index and the reionization redshift.

In appendix \ref{app:fisher} we review the the Fisher matrix error estimation procedure  in the case of a CMB experiment. 
In the application here  we use the modified CAMB version used in \S \ref{s:mcmc} to compute the $C_{\ell}$. 

Beyond error estimations and forecasting, there exists another powerful tool encoded in the Fisher formalism. This enables one, without recomputing the full covariance matrix, to calculate the shift in the best fit parameters $\delta\theta_\alpha$ due to the fact that an arbitrary number of parameters has been fixed to a wrong fiducial value. See \cite{Heavens-Kitching-Verde07} for more details. 
 
In general, given a number $p$ of parameters  $\Psi_\gamma \; (\gamma=1,..,p)$ fixed to an incorrect value, which differs from the "true" value for an amount $\delta \Psi_\gamma$, the  resulting shift on the  best fit value of the other $n$ parameters $ \theta_\alpha \; (\alpha=1,...,n)$ is:
 \be
\delta \theta_\alpha = - \left(F^{-1}\right)_{\alpha \beta} \, S_{\beta  \gamma} \delta \Psi_\gamma.
\ee

\begin{table*}
\begin{center}
\begin{tabular}{|cccccc|} 
\hline
\hline
Channels (in GHz)&44&70&100&143&217\\
Beam FWHM &24&14&10&7.1&5\\
Final noise per arcminute $\mu$K &180.36&180.74&77.07&57.46&94.42\\
\hline
\end{tabular}\label{tab:exp-setting}
\caption{Planck experiment setting (upper table) and experimental specifications for the surveys used in this paper (lower table): redshift range $z$, survey area (A) in squared degrees, fraction of the sky $f_{sky}$ and mean galaxy number density $\bar n$
}
\begin{tabular}{|ccc|} 
\hline
$\;$&BOSS&EUCLID\\
\hline
$z$& $<0.7$ & $<2$ \\
A ($deg^2$)&8000&20000\\
$f_{sky}$&0.2&0.8\\
 $\bar n$&2.66 $\,10^{-4}$&1.53 $\,10^{-5}$\\
\hline
\end{tabular}
\end{center}
\end{table*}

Here $\left(F^{-1}\right)_{\alpha \beta}$ is the sub-matrix of the  inverse Fisher matrix,  corresponding to the $\theta_\alpha$ parameters (i.e. the inverse of the Fisher matrix, without the rows and columns corresponding to the ``incorrect" parameters) and $S_{\beta  \gamma}$ is the Fisher sub-matrix including also the $\Psi_\gamma$ parameters.

Here in particular we are interested in computing the shift in the best fit parameters induced by setting the amount of isocurvature $f_{iso}$ to an incorrect value: 
$\delta \Psi_\gamma=\delta f_{iso}$ and $\theta_\alpha=\{r, n_s, d n_s/d\ln k, z_{re}, \omega_b, \omega_c, h, A_s\}$.


Moreover we want an estimate of this effect on the value of the sound horizon at radiation drag $r_s(z_d)$, which is a parameter not directly  included in the Fisher matrix.  This can be done via a change of base  in the Fisher approach.
We calculate $r_s(z_d)$ (see Eq. (\ref{eq:rsz}) numerically with values for the cosmological parameters given by Eq. (\ref{eq:fisher-params}). 
In general, if $p=(p_1,...,p_N)$ are the parameters in the Fisher matrix $F^{old}_{n m}$ and $q=(q_1,...,q_Q)$ are the new ones, the new Fisher matrix $F^{new}_{i j}$ respect to these will be:
\be
F^{new}_{i j} = \frac{\partial p_n}{\partial q_i} F^{old}_{n m} \frac{\partial p_m}{\partial q_j}.
\ee
In our case the Jacobian matrix $J_{ni}= \frac{\partial p_n}{\partial q_i}$ will have non-zero relevant terms corresponding to: $\frac{\partial h}{\partial r_s}$, 
$\frac{\partial \omega_b}{\partial r_s}$ and $\frac{\partial \omega_c}{\partial r_s}$. 

The result are shown in the table \ref{t:fisher}: the second and third columns refer, respectively, to the fiducial values for the parameters and their $1-\sigma$ errors calculated from the Fisher matrix, while in the third column we report the shifts on parameters in the form of $\partial \theta_{\alpha}/\partial f_{iso}$,  that is the shift on the parameters per unit shift in $f_{iso}$. The parameters more affected (labeled with $\blacktriangleright$ in the table) are the scalar amplitude and spectral index $A_s, \,n_s$, the Hubble constant, parameterized by  $h$, the cold dark matter physical density $\omega_c$ and, as a consequence,  the value of the sound horizon at radiation drag $r_s(z_d)$.
The fiducial value for the amount of isocurvature used in the Fisher calculation is $f_{iso}=-0.01$ so, in the case that the universe were adiabatic, we would have a shift on parameters corresponding to $\delta f_{iso}=+0.01$. In this case it is interesting to compare the effect on the parameters $\theta_\alpha$ induced by such a (small) shift to the expected error on the parameters: the values of this ratio (\%) are shown in the last column of table \ref{t:fisher}. 
The shift on the parameters, for a $\delta f_{iso}=0.01$, turns out to be bigger than the error ($\simeq$ 120\% of $\sigma_\alpha$) for  $r_s(z_d)$, $h$ and $\omega_c$ and even more for $n_s$ ($\simeq$ 150\%). The amount of the shift in the cosmological parameters due to the presence of an isocurvature fraction found with this Fisher analysis is in  agreement with what we found  with the previous MCMC analysis,  even if the base parameter set is different in the two applications (in the MCMC approach $r$ and $d n_s/d\ln k$ were not left as parameters). In particular, we find in both case the same degeneracy direction for the parameters $H_0$ and $r_s(z_d)$ with $f_{iso}$. 
\begin{table*} 
\caption{Fisher matrix for a Planck-like set-up - Shift on parameters due to the fact that $f_{iso}$ is fixed to an incorrect value ($\delta f_{iso} \neq 0$). The fiducial value adopted here is $f_{iso}=-0.01$.
	}
\begin{center}
\begin{tabular}{|ccccc|} 
\hline
\hline
Parameter $\alpha$ &Fid. Value&Fisher error $\sigma_\alpha$&Shift  $\frac{\delta\theta_\alpha}{\delta f_{iso}}$&$|\frac{\delta\theta_\alpha}{\sigma_\alpha}|_{\delta f_{iso}=0.01}(\%)$\\
\hline
$r$&$0.01$&$0.02299$&$0.1563$&6.79\\             
$\blacktriangleright n_s$&$0.963$&$0.00433$&$0.6612$&152.7\\            
$d n_s/ d\ln k$&$0.0$&$0.0053$&$-0.2292$&43.24\\        
$z_*$&$0.840$&$0.0077$&$-0.0196$&0.25\\    
$\omega_b$&$0.02273$&$0.000127$&$-0.0005$&3.93\\                        
$\blacktriangleright  \omega_c$&$0.1099$&$0.00115$&$-0.139$&120.55\\                           
$\blacktriangleright  h$&$0.72$&$0.00581$&$0.71467$&123.0\\                
$\blacktriangleright  A_s$&$0.8169$&$0.00825$&$0.35752$&43.33\\                     
$f_{iso}$&$-0.01$&$0.01123$&$/$&/\\
$\blacktriangleright r_s(z_d) \; Mpc$&$149.641$&$ 0.359$&$-44.215765$&123.19\\ 
\hline
\end{tabular}\label{t:fisher}
\end{center}
\end{table*}

While the $n_s$ shift is large, we expect the amplitude of this effect to be specific of the model we have considered ($n_{iso}=n_{ad}$) and thus the  amplitude of this shift to depend strongly on the choice adopted for the relation between the two spectral slope indices.  The quantitative effect on $A_s$ can also be affected by this choice. 
However, the effect on the other parameters such as  $r_s(z_d)$, is expected to be less sensitive, at least qualitatively, to this choice as the signal comes from a more localized region in multipoles $\ell$ (see discussion in \cite{Kurki05}). 
Thus while the effect on $n_s$ may be relevant to analyses geared towards determining the shape of the primordial power spectrum and investigating the implications for inflation, here we concentrate on the implications of a systematic shift on $r_s(z_d)$ and  parameters  yielding the Universe's expansion history.

\begin{figure*}
\includegraphics[width=17.3cm,height=8.3cm]{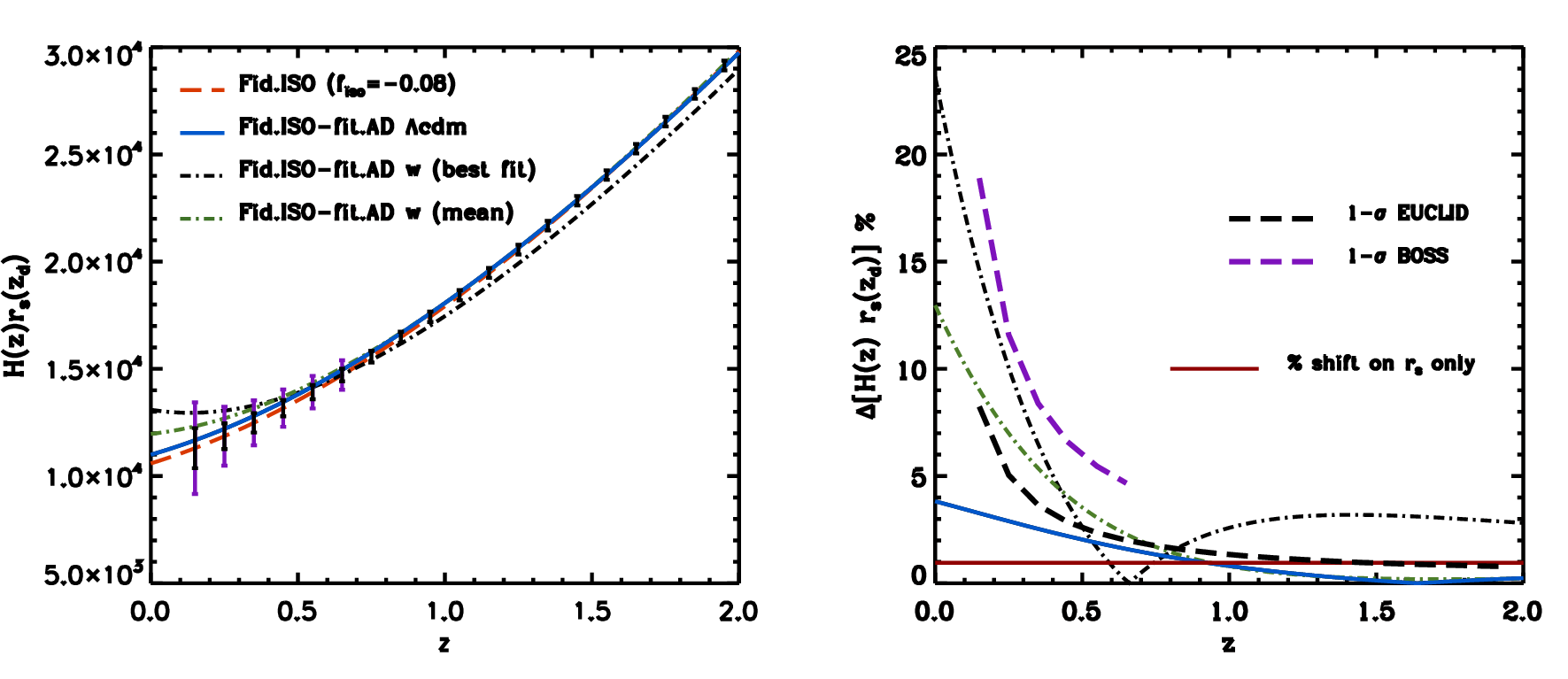}%
\caption{
BAO observable  $H(z) \, r_s(z_d) $ and isocurvature (fiducial curvaton model with mixed ad+cdi modes $f_{iso}=-0.08$ ): LEFT panel: the dashed orange line shows   $H(z) \, r_s(z_d) $ %
 as a function of the redshift $z$ for the fiducial curvaton model, the blue line for the fiducial curvaton model fitted with an adiabatic model (Fid.ISO-fit.AD) for a standard $\Lambda$CDM cosmology and the dot-dashed line for the same model %
but for a cosmology with varying dark energy equation of state w [the black line refers to the curve obtained by using the best fit parameters while the green curve by using the mean values for the parameters].
The purple error bars are the forecasted $1-\sigma$ errors for the BOSS survey \cite{Boss}, while the black ones are for the EUCLID survey \cite{Euclid}.
The RIGHT panel plot shows the percentile variation of $H(z) \, r_s(z_d)$ %
 calculated by comparing the value of this observable as a 
function of $z$ for the underlying fiducial curvaton model with the fiducial curvaton model fitted with an adiabatic model (Fid.ISO-fit.AD) for a standard $\Lambda$CDM cosmology (blue line) and with the same model but with varying w (dot-dashed lines). The dashed lines refer to the percent  $1-\sigma$ errors on $H(z) \, r_s(z_d) $ for the BOSS survey (purple) and EUCLID (black). The red straight line refers to the percentile variation of $H(z) \, r_s(z_d)$ when only $r_s(z_d)$ is shifted because of the presence of the isocurvature (Fid.ISO vs Fid.ISO-fit.AD $\Lambda$CDM, see section \ref{rszd}).
 }
\label{fig:hzrs}
\end{figure*}

\section{Isocurvature and BAO: Future surveys forecast and implication for DE}\label{sec:results}

In  light of the findings obtained so far, the aim of this section is to explore the effect on the BAO observables of allowing for an isocurvature fraction. 
 The BAO observables are  Eqs. (\ref{e:bao-par}) and (\ref{e:bao-per}): $B_{\parallel}(z)= H(z) \,r_s(z_d)$ and $B_{\perp}(z)=D_A(z)/r_s(z_d)$. 
These are the quantities,  respectively parallel and transverse to the line-of-sight, that future galaxy surveys like BOSS \cite{Boss} and EUCLID \cite{Euclid} plan to measure with high accuracy. We quantify the effect that the presence of isocurvature could introduce on the  expected values  and compare it with  forecasted errors on such quantities.  In order to do so we use the code and the method described in \cite{Seo-Eisenstein}. The experimental specifications for both surveys are listed in table \ref{tab:exp-setting}.
We refer to section \ref{s:mcmc} for the calculation of the expansion history $H(z)$ and the sound horizon at radiation drag $r_s(z_d)$.  Similarly, we  compute the  co-moving angular diameter distance $D_A(z)$ from  Eq. \ref{eq:ang-dist}. 

\begin{figure*}
\includegraphics[width=17.3cm,height=8.3cm]{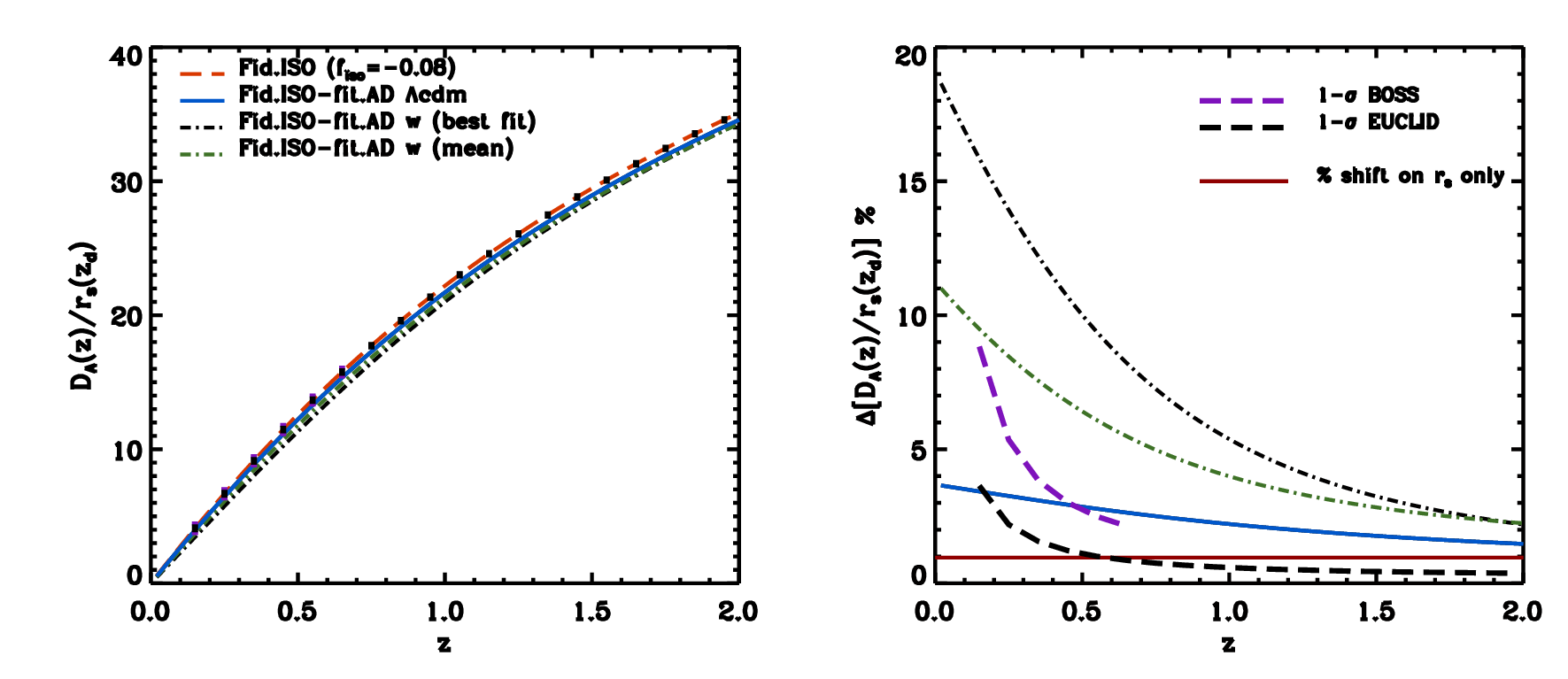} 
\caption{BAO observable $\frac{D_A(z)}{r_s(z_d)}$ and isocurvature (mixed ad+cdi model): LEFT panel: %
the dashed orange line shows $\frac{D_A(z)}{r_s(z_d)}$ %
as a function of the redshift $z$ for the fiducial curvaton model, the blue line for the fiducial curvaton model fitted with an adiabatic model (Fid.ISO-fit.AD) for a standard $\Lambda$CDM cosmology and  the dot-dashed line for the same model %
but for a cosmology with varying dark energy equation of state w [the black line refers to the curve obtained by using the best fit parameters while the green curve by using the mean values for the parameters].
The purple error bars are the forecasted $1-\sigma$ errors for the BOSS survey \cite{Boss}, while the black ones are for the EUCLID survey \cite{Euclid}.
The RIGHT panel plot shows the percentile variation of $\frac{D_A(z)}{r_s(z_d)}$ 
 calculated by comparing the value of this observable as a 
function of $z$ for the underlying fiducial curvaton model with the fiducial curvaton model fitted with an adiabatic model (Fid.ISO-fit.AD) for a standard $\Lambda$CDM cosmology (blue line) and with the same model but with varying w (dot-dashed lines) where the black line refers to the curve obtained by using the best fit parameters while the green curve by using the mean values for the parameters. The dashed lines refer to the percent  $1-\sigma$ errors on $\frac{D_A(z)}{r_s(z_d)}$ for the BOSS survey (purple) and EUCLID (black).
The red straight line refers to the percentile variation of $D_A(z)/r_s(z_d)$ when only $r_s(z_d)$ is shifted because of the presence of the isocurvature (Fid.ISO vs Fid.ISO-fit.AD $\Lambda$CDM, see section \ref{rszd}).
}
\label{fig:dazrs}
\end{figure*}

The results are summarized in Fig. \ref{fig:hzrs}  and Fig. \ref{fig:dazrs}. On the left panel we show the redshift dependence of the observable quantity ($r_s(z_d)H(z)$ and $D_a(z)/r_s(z_d)$ respectively) and the forecasted error-bars for  forthcoming (BOSS) and planned (Euclid) surveys.
On the right panel we show the \% shift of the quantity (thin lines)  and the forecasted errors (dashed thick lines, black and purple).  We should distinguish two effects.

 The first is  a systematic effect arising from the shift on the CMB-based $r_s(z_d)$ determination. This is shown as the red solid  line at the 1\% level for a $\Delta f_{\rm iso}=0.08$. This effect  is negligible for a survey like BOSS but it is at the level of statistical errors  for the Euclid $H(z)$ over 7 or 8 redshift bins and  at or above Euclid statistical errors on $D_A(z)$ over  more than 10 bins. This indicates that the presence of an isocurvature component at a level well within the  allowed   range, if  neglected, can introduce a systematic error in the interpretation of the BAO signal that is  comparable if not larger than the statistical errors. 
Similarly, an isocurvature component,  $f_{iso}$, of a magnitude allowed at  $3-\sigma$-level by Planck data could yield a $1-\sigma$ systematic effect on $B_{\parallel}$ and therefore $H(z)$ and   an  $f_{iso}$ of a magnitude allowed at  $2.5-\sigma$-level could yield a  $1-\sigma$ systematic  shift  on $B_{\perp}$ and thus $D_A(z)$.

One should keep in mind that a $1\%$ shift on $H(z)$ at $z\sim 1$ implies a shift of $0.05$ on $w_0$ and that a 1\% shift on $D_A$ at $z\sim 1$  implies a shift of $0.07$ on $w_0$.

The second effect is a mis-match of the expansion history as inferred from CMB and as measured by BAO surveys. The thin lines in  Figs. \ref{fig:hzrs},  \ref{fig:dazrs} show the CMB-predicted BAO observables (which combine the shift on the expansion history and the shift on the sound horizon at radiation drag).  The solid blue line corresponds to the Fid.ISO-fit.AD case. The dot-dashed lines are the  best fit (black) and mean value (green) for the Fid.ISO-fit.ADw case. The mis-match  is clearly above the BAO surveys statistical errors over a wide range in redshift for the transversal measurement for BOSS and for both, radial and transversal  measurements for Euclid.
The mis-match between CMB predicted and the measured  expansion histories has been proposed  as a signature for deviations from a $\Lambda$CDM cosmology in the form of deviations from Einstein's gravity  e.g.,\cite{AcquavivaVerde07, Ishak06}, couplings in the dark sector e.g.,\cite{LRMVJ} or time evolving dark energy. Here we  add another possibility to this list: an incorrect assumption about the nature of primordial perturbations.

We should note however that  including $f_{iso}$ as a parameter  in the CMB analysis solves this issue.  Should there be a small isocurvature contribution, adding the $f_{iso}$ parameter to the analysis will recover correctly $r_s$ and the expansion history  while  the  error-bars  degrade by quarter of a sigma.

\section{Conclusions}\label{sec:conclusions} 

In this paper we have investigated how a biased determination of of the sound horizon due to a incorrect assumption about the nature of the initial  perturbations affects cosmological parameters measurements from future BAO data.
It is important to keep in mind that analyzing the CMB data with the prior assumption of purely adiabatic 
initial conditions when the real universe contains even a small isocurvature contribution, could lead to an incorrect 
determination of the cosmological parameters. In fact, the presence of an isocurvature component changes the shape 
and the location of the CMB acoustic peaks, mimicking the effect of parameters such as $\Omega_mh^2$ , $H_0$ and $w$. Here  concentrated on  wether such an effect has an impact on Òstandard rulersÕ, like the sound horizon at recombination,  inferred from CMB observations. 
This issue turns out to be very delicate and important especially in view of the next generation of galaxy surveys 
which aims at probing with high accuracy the late time expansion and thus the nature of dark energy by means of 
BAO at low redshift ($z < 2$). 

We find that the presence of an  isocurvature contribution, of a magnitude still allowed by present data, can affect both the size of the sound horizon at radiation drag and the CMB-inferred expansion history, in a non-negligible way.  For our forecasts we used an experiment with the characteristics of  the Planck  mission.
We find that the  systematic error introduced  in the  sound horizon at radiation drag propagates into a systematic error on quantities like $H(z)$ and $D_A(z)$ that  can be  comparable or larger than the statistical error over a wide range of redshift bins for  future  BAO experiments.

A crucial assumption of the BAO technique  is the possibility to measure accurately and robustly the acoustic 
sound horizon at radiation drag. Generally CMB observations yield a robust determination of 
such standard ruler although there could be possible systematic effects introduced by deviations from the standard 
evolution of the early universe.  In this paper we have presented another 
possible deviation from the minimal, standard cosmological model that could affect the BAO interpretation as standard 
rulers if not taken into account: the presence of an isocurvature contribution to the initial conditions for primordial 
perturbations. 

In addition  the neglected presence of an isocurvature component in the initial conditions  introduces a mis-match between the expansion history as inferred from CMB and as measured by BAO  or Supernovae surveys. The mis-match is  above the BAO surveys statistical errors over a wide range in redshift for both on-going and future surveys. Such a high redshift-low redshift  mismatch between CMB predicted and the 
measured expansion histories has been proposed as a signature for deviations from a $\Lambda$CDM cosmology in the form  of deviations from EinsteinÕs gravity, couplings in the dark sector, or time evolving dark energy.  Here we add another possibility to this list, an incorrect assumption about  the nature of primordial perturbations. 

 In the paper we have  used the curvaton scenario as a working example for a model that gives rise to a small 
fraction of correlated isocurvature perturbations.  The curvaton has $n_{ad} = n_{iso}$, therefore our findings are quantitative only for this case, but hold qualitatively for other models.

In order to recover an unbiased determination  for  the sound horizon and dark energy parameters, a component of isocurvature perturbations must be included in the model  when analyzing CMB data.  Fortunately, doing so does not increase parameter errors significantly.

\section*{Acknoweldgments}
AM is supported by a MICINN FPI grant BES-2009-013471. LV is supported by FP7-PEOPLE-2007-4-3 IRG n 202182, FP7 IDEAS-Phys.LSS.240117 and  by MICINN grant  AYA2008-03531.  MB is supported by the Department of Energy. AM thanks N.Bartolo,  B. Reid  and D. Eisenstein for stimulating discussions.
\appendix

\section{BAO and Isocurvature}

The baryon acoustic oscillations (BAO) responsible for the acoustic peaks structure imprinted in the CMB are predicted to
be present in the late-time clustering of galaxies as a series
of weak modulations in the amplitude of fluctuations
as a function of scale \cite{Seo-Eisenstein}.

In general, the line-of-sight ($r_{\parallel}$) 
and transverse ($r_{\perp}$)
co-moving sizes of an object or a feature at redshift $z$ are related to the correspondent observed sizes $\Delta z$ (redshift slice) and $\Delta \theta$ (angular separation) 
by the expansion history $H(z)$ and the angular diameter distance $D_A(z)$ through: $r_{\parallel}=\frac{c \Delta z}{H(z)}$ and $r_{\perp}= D_A(z) \Delta \theta$,
being the co-moving angular diameter distance: %
\be\label{eq:ang-dist}
D_A(z)=c \int_0^z \frac{dz'}{H(z')}   %
\ee
and
\be\label{eq:hz}
H(z)=H_0 \,\sqrt{\sum_i \Omega_i (1+z)^{3(1+w_i)}}
\ee
where $H_0$ is the Hubble parameter at the present time and $\Omega_i$ and $w_i$ are, respectively, the density parameter and the equation of state parameter of a given species $i$.

The key parameter associated to the BAO in the primordial photon-baryon fluid, which leave an imprint in the large scale matter distribution, is the size of the sound horizon at the radiation drag redshift $z_d$, i.e. at the time when baryons were realized from the photons.
\be\label{eq:rsz}
r_s(z_d)= \frac{1}{H_0} \int^{\infty}_{z_d} \frac{dz}{E(z)} c_s(z) 
\ee
with the sound speed:
\be
c_s(z)=\frac{c}{\sqrt{3(1+ \frac{3}{4} \frac{\Omega_{b}}{\Omega_{\gamma}} \frac{1}{1+z}})},
\ee
being $\Omega_b$ and $\Omega_\gamma$, respectively, the baryon and photon density parameters at the present time. 
The function
$E(z)=(\sum_i \Omega_i (1+z)^{3(1+w_i)})^{\frac{1}{2}}$ %
up to the time of radiation drag accounts for the contribution of radiation (photons and neutrinos) and matter (baryons and cold dark matter) so that takes the form: %
\be
E(z) \simeq \sqrt{\Omega_r (1+z)^4+ \Omega_m(1+z)^3},
\ee

where $\Omega_r=\Omega_{\gamma_0}(1+0.2271 N_{\rm eff})$ is the radiation term, $\Omega_{\gamma_0}$ is the present photon density parameter and $N_{\rm eff}$ the neutrino effective number that we set to the standard choice of 3 neutrino species.

The
CMB could in principle provide such a 'standard ruler', %
 so that, by measuring the angle subtended by this ruler as a function of redshift we can map out the angular
diameter distance $D_A(z)$ and  by measuring the redshift interval associated
with this distance we can map out the Hubble parameter
$H(z)$. This is the key idea on which the next generation of future BAO galaxy surveys is based. %

We will thus be interested in the observables, respectively along and perpendicular to the observer line-of-sight:
\be\label{e:bao-par}
B_{\parallel}(z)= H(z) \,r_s(z_d)
\ee
and
\be\label{e:bao-per}
B_{\perp}(z)=\frac{D_A(z)}{r_s(z_d)}.
\ee

In view of high precision data it is mandatory to test how robust is the method and the determination of the standard ruler. In particular it will be important to know/account for all the possible systematics that could affect such measurement.
As already pointed out in the Introduction, relaxing the hypothesis of adiabaticity for the initial conditions compromises the precision of  parameter estimation:  for example, in the case of the sound horizon at radiation drag $ r_s(z_d)$, the presence of an isocurvature contribution does not change the dependence on the cosmological parameters, %
but it will produce a difference in the measured value with respect to the one obtained assuming a purely adiabatic model. Analogously the presence of an isocurvature contribution would produce an "adjustment" in the cosmological parameters which could lead to a wrong interpretation of the expansion history parameter.

\section{Fisher matrix for CMB}\label{app:fisher}

The Fisher matrix is defined as \cite{Fisher}:
\be
F_{ij}=- \langle \frac{\partial^2 \text{ln}{\cal L}}{\partial \theta_i \partial \theta_j}\rangle,
\ee
where ${\cal L}$ is the (gaussian) likelihood for a set of parameters {\bf$\theta$}:
\be
{\cal L}=\frac{1}{(2 \pi)^{n/2}| \text{det} {\bf Cov}|^{1/2}} \text{exp} \left[ - \frac{1}{2} \sum_{ij} (D-y)_i  {\bf Cov}^{-1}(D-y)_j \right].
\ee
$D$ and $y \equiv y(\theta)$ are respectively the data set and the fiducial theoretical model and ${\bf Cov}$ is the covariance matrix: ${\bf Cov}_{ij}= \langle(D_i-y_i)(D_j - y_j) \rangle$. 

In general  the estimate of the covariance for the parameters is given by: $\sigma_{ij}^2 \geq (F^{-1})_{ij}$ and the marginalized error (i.e. calculated with the full covariance matrix) on a given parameter $\theta_i$ is:
\be
\sigma_{\theta_i} \geq \sqrt{(F^{-1})_{ii}}.
\ee
 So, once computed the covariance matrix and having a fiducial model with a known dependence on the parameters, the Fisher matrix gives the expected errors. This is a very powerful tool extremely useful in experiments design which enables to forecast the parameter errors before doing the experiment. 
 
 In the case of a full sky and noiseless CMB experiment the gaussianly distributed signal in the sky can be described by the vector $\langle a_{\ell m}^T, a_{\ell m}^E , a_{\ell m}^B \rangle$, with $a_{\ell m}^X, \; X=T,E,B$ the spherical harmonic coefficients for the temperature, the E and the B polarization modes. 
In this case the Fisher matrix will take the form:
\be
F_{ij}^{CMB}=\sum_{XY} \sum_\ell \frac{\partial C^X_\ell}{\partial \theta_i} (\mathcal M_\ell^{XY})^{-1} \frac{\partial C^Y_\ell}{\partial \theta_j}
\ee
where $C_\ell$ is the CMB angular power spectrum: $\langle a_{\ell m} a^*_{\ell' m'} \rangle = \delta_{\ell \ell'} \delta_{m m'} C_\ell$ and $\mathcal M_\ell^{XY}$, $X,Y=TT,EE,TE,BB,...$ are the elements of the matrix:
\begin{equation*}
{\mathcal M}_\ell=\frac{2}{2 \ell +1}
\left(
    \begin{array}{cccc}
(C^{TT}_\ell)^2&(C^{TE}_\ell)^2&C^{TT}_\ell C^{TE}_\ell&0\\
(C^{TE}_\ell)^2&(C^{EE}_\ell)^2&C^{EE}_\ell C^{TE}_\ell&0\\
C^{TT}_\ell C^{TE}_\ell&C^{EE}_\ell C^{TE}_\ell&1/2[(C^{TE}_\ell)^2+C^{TT}_\ell C^{TE}_\ell]&0\\
0&0&0&(C_\ell^{BB})^2\\
\end{array}
\right)
\,\,,
\end{equation*}
which takes into account the correlation between TT, EE and TE.

\section{Isocurvature modes and the CMB: parameter degeneratios}\label{sec:iso-cmb}

The main effect of pure isocurvature perturbations on the CMB is a $\frac{\pi}{2}$ phase shift on the acoustic oscillation with respect to the adiabatic case which moves the peak structure in the CMB temperature spectrum to smaller scales. This could be qualitatively described as a phenomena of 'compensation' \cite{Hu-Sper-Whi97}. Any source of isocurvature density perturbations must be compensated to keep the total density fluctuations small, this means that there must an anticorrelation between the isocurvature source and the photon density perturbations (being the photons the dominant dynamical component at this stage).

More specifically, in the adiabatic case the photon-baryon fluid begins to compress itself due to its self-gravity. As pressure tries to stop the compression,
the potential decays. The fluid is left in a highly compressed state and self-gravity acts to enhance the first compression of a cos(ks) series. Peaks occur at:
\be
\ell^{AD}_m=m \,  \ell_A  \;\;\;\;  m=0,1,2,...
\ee
where $\ell_A$ is the ratio between the co-moving  angular diameter distance to the last scattering surface $D_A(z*)$ and the sound horizon at last scattering $r_s(z*)$ (Eqs. (\ref{eq:ang-dist}) and (\ref{eq:rsz})):
\be
\ell_A= \pi \frac{D_A(z*)}{r_s(z*)}.
\ee
 Baryon
drag enhances all odd peaks. 
 
 In the isocurvature case the potential fluctuations vanish initially and then grow in anticorrelation with the
photon fluctuations until horizon crossing. Because of photon pressure near horizon crossing the compensation mechanism  that keeps the curvature vanishing breaks letting
the potential fluctuation grow from zero. To compensate this the photons rarefy inside the potential wells. At horizon crossing the photon pressure resists the accompanying rarefaction and the fluid
turns around and begins falling into the potential wells. The self-gravity of the
photon-baryon fluid drives the sine rather than
the cosine oscillation. 
The position of the acoustic peaks is given by: 
\be
\ell^{iso}_m = (m - \frac{1}{2})\, \ell_A 
\ee
and baryon drag enhances all even peaks.

Therefore, accounting for an isocurvature contribution induces a shift in the position of the CMB peaks to smaller scales: the more isocurvature fraction, bigger the peak displacement. However, this effect can be compensated by a change in other cosmological parameters, namely the Hubble parameter $H_0$. Qualitatively an enhancement in $H_0$ yields in fact a shift of the peaks position towards larger scales while a decrement results to a shift to smaller scales which can thus mimic or compensate the presence of isocurvature modes. For this reason the two parameters $f_{iso}$ and $H_0$ are said to be degenerate. For a more detailed discussion on isocurvature and degenerations with the CMB parameters, in particular for the more complicated cases of the mixture of adiabatic and isocurvature modes, see for example: \cite{Langlois00,Kurki05,Trotta01}.



\begin{thebibliography}{10}
\bibitem{mukhanov-adiabatic}
V.~ Mukhanov, H.A. ~Feldman and R.H.~ Brandenberger,
\newblock Phys. Rept. {\bf 15}, 203 (1992).
\bibitem{Brandenberger92} 
Brandenberger, R., Mukhanov, V., \& Prokopec, T.\ 
1992, Physical Review Letters, 69, 3606 

\bibitem{KomatsuWMAP7} 
Komatsu, E., et al.\ 
2010, arXiv:1001.4538


\bibitem{Linde1985}
A.~D.~Linde,
Phys.\ Lett.\ B {\bf 158}, 375 (1985);
A.~D.~Linde and V.~Mukhanov,
Phys.\ Rev.\ D {\bf 56}, 535 (1997);
\bibitem{Kofman1986}
L.~A.~Kofman and A.~D.~Linde,
Nucl.\ Phys.\ B {\bf 282}, 555 (1987);

\bibitem{Mollerach1990}
S.~Mollerach,
Phys.\ Lett.\ B {\bf 242}, 158 (1990);

\bibitem{Peebles1999}
P.~J.~E.~Peebles,
Astrophys.\ J. {\bf 510}, 523 (1999).

\bibitem{Polarski1994}
D.~Polarski and A.~A.~Starobinsky,
Phys.\ Rev.\ D {\bf 50}, 6123 (1994);
M.~Sasaki and E.~D.~Stewart,
Prog.\ Theor.\ Phys.\  {\bf 95}, 71 (1996);
M.~Sasaki and T.~Tanaka,
Prog.\ Theor.\ Phys.\  {\bf 99}, 763 (1998).

\bibitem{Langlois99}
 Langlois, D.,
   PhysRevD.59, 123512 (1999)
  

\bibitem{Bartolo2001}
N.~Bartolo, S.~Matarrese and A.~Riotto,
Phys.\ Rev.\ D {\bf 64}, 123504 (2001); 
C.~Gordon, D.~Wands, B.~A.~Bassett and R.~Maartens,
Phys.\ Rev.\ D {\bf 63}, 023506 (2001);
D.~Wands, N.~Bartolo, S.~Matarrese and A.~Riotto,
Phys.\ Rev.\ D {\bf 66}, 043520 (2002).
J.~Garc\'{\i}a-Bellido and D.~Wands,
Phys.\ Rev.\ D {\bf 53}, 5437 (1996); {\bf 52}, 6739 (1995).


\bibitem{Bucher-general-2001}
M.~Bucher\&al. 
\newblock Phys. Rev. D {\bf 62}, 083508 (2000)


\bibitem{Axenides:1983hj}
  M.~Axenides, R.~H.~Brandenberger and M.~S.~Turner,
  Phys.\ Lett.\ B {\bf 126} (1983) 178.


\bibitem{Lindeaxion} 
  A.~D.~Linde, 
  JETP Lett.\  {\bf 40} (1984) 1333 
  [Pisma Zh.\ Eksp.\ Teor.\ Fiz.\  {\bf 40} (1984) 496]; 
  Phys.\ Lett.\ B {\bf 158}, 375 (1985); 
  Phys.\ Lett.\ B {\bf 201} (1988) 437. 
 
\bibitem{SeckelTurner}
  D.~Seckel and M.~S.~Turner,
  Phys.\ Rev.\ D {\bf 32} (1985) 3178. 

\bibitem{hybrid}
  A.~D.~Linde, 
  Phys.\ Lett.\ B {\bf 259} (1991) 38.
   
\bibitem{TurnerWilczek} 
  M.~S.~Turner and F.~Wilczek, 
  Phys.\ Rev.\ Lett.\  {\bf 66} (1991) 5. 

\bibitem{LindeLyth}
 A.~D.~Linde and D.~H.~Lyth,
 Phys.\ Lett.\ B {\bf 246} (1990) 353.

\bibitem{Lyth:1991}
  D.~H.~Lyth,
  Phys.\ Rev.\ D {\bf 45} (1992) 3394.
\bibitem{Shellard:1997}
  E.~P.~S.~Shellard and R.~A.~Battye,
  arXiv:astro-ph/9802216.
  
\bibitem{Kawasaki1995axion}
M.~Kawasaki, N.~Sugiyama and T.~Yanagida,
Phys.\ Rev.\ D {\bf 54}, 2442 (1996);

\bibitem{curvaton1}
A. D. ~ Linde and V. F. ~ Mukhanov,
\newblock Phys. Rev. D{\bf 56}, 535 (1997), astro-ph/9610219

\bibitem{curvaton2}
D.~H. Lyth and  D.~Wands,
\newblock Phys. Lett.B {\bf 524}, 5-14 (2002).

\bibitem{curvaton3}
D.~H. Lyth, C.~Ungarelli, and D.~Wands,
\newblock Physical Review D {\bf 67}, 023503 (2003).




\bibitem{Efstathiou86}
G.~Efstathiou and J.R.~Bond,
\newblock MNRS {\bf 218}, 103 (1986)

\bibitem{Page03}
L.~Page et al.
\newblock Astrophys. J. Suppl. {\bf 148} 233 (2003) 


\bibitem{Hinshaw06}
G.~Hinshaw et al.
\newblock Astrophys. J. Suppl.{\bf 170} 288 (2007)

\bibitem{Enqvist02}
K.~Enqvist, H.~Kurki-Suonio and J.~Valiviita,
\newblock  Phys. Rev. D {\bf 65}, 043002 (2002)

\bibitem{Beltran-04}
Maria Beltran, Juan Garcia-Bellido, Julien Lesgourgues, Alain Riazuelo
Journal-ref: Phys.Rev. D70 (2004) 103530 

\bibitem{Beltran-05}
Maria Beltran, Juan Garcia-Bellido, Julien Lesgourgues, Matteo Viel
Phys.Rev.D72:103515,2005 

\bibitem{Valiviita09}
J.~Valiviita and T.~Giannantonio,
\newblock  Phys. Rev. D {\bf 80}, 123516 (2009)

\bibitem{Dunkley05}
J.~Dunkley et al.,
Phys. Rev. Lett.  {\bf 95}, 261303 (2005)

\bibitem{Keskitalo07} 
Keskitalo, R., Kurki-Suonio, H., Muhonen, V., Valiviita, 
J.\ 2007, Journal of Cosmology and Astro-Particle Physics, 9, 8

\bibitem{CAMB}
http://camb.info

\bibitem{Kurki05}
H.Kurki-Suonio, V.~Muhonen and J.~Valiviita
\newblock Phys. Rev. D {\bf 71}, 063005 (2005)


\bibitem{Valiviita03}
J.~Valiviita and V.~Muhonen,
\newblock  Phys. Rev. Lett. {\bf 91}, 131302 (2003)

\bibitem{trotta2003} 
R.~Trotta
New Astron. Rev. {\bf 47}, 769-774 (2003)


\bibitem{Langlois00}
D.~Langlois and A.~Riazuelo
\newblock Phys. Rev. D {\bf 62}, 043504 (2000)

\bibitem{Bucher00}
M.~Bucher, K.~Moodley and N.~Turok,
Cosmology and Particle Physics, {\bf 555}, 313 (2001)

\bibitem{Sollom09} 
Sollom, I., Challinor,A., \& Hobson, M.~P.\ 2009, \prd, 79, 123521

\bibitem{Bean2006}
R.~Bean, J.~Dunkley and E.~Pierpaoli,
Phys.Rev. D {\bf 95}  063503 (2006)

\bibitem{Trotta01}
R.~Trotta, A.~Riazuelo and R.~Durrer
Phys. Rev. Lett.  {\bf 87}, 231301 (2001)

\bibitem{bucher-pol}
M.~Bucher, K.~Moodley and N.~Turok,
\newblock Physical Review Lett {\bf 87}, 191301 (2001).


\bibitem{zd-fit-eis-hu}
 D. J.~Eisenstein and W.~Hu,
 \newblock ApJ, {\bf 496}, 605 (1998)
 
\bibitem{Seo-Eisenstein}
H. J. ~Seo and D. J.~Eisenstein,
\newblock ApJ, {\bf 598}, 720 (2003)


\bibitem{EisensteinWhite04}
D. J.~Eisenstein and M.~White,
Phys.Rev.D {\bf 70}, 103523 (2004)

\bibitem{deBernardis09}
F.~De Bernardis et. al.,
Phys. Rev. D {\bf 79}, 043503 (2009)

\bibitem{Linder08}
E.~Linder and G.~Robbers,
JCAP 0806, 004 (2008)


\bibitem{Hu-Sper-Whi97}
W.~Hu, D.N.~Spergel and M.~White,
\newblock Phys.Rev. D {\bf 55}, 3288-3302 (1997)



\bibitem{Peiris03}
H.V.~Peiris et al.,
\newblock Astophys. J. Suppl. {\bf 148}, 213 (2003)


\bibitem{constant-cur}
J.M.~ Bardeen, P.J. ~Steinhardt and M.S. ~Turner, 
Phys. Rev. D {\bf 28}, 679 (1983) 
and
D.H. ~Lyth,  
\newblock Phys. Rev. D {\bf 31}, 1792 (1985).

%
\bibitem{bluebook}
http://www.rssd.esa.int/index.php?project=PLANCK\&page=index

\bibitem{Baumannetal}
D.Baumann et al.,
AIP Conf. Proc. 1141, 10-120 (2009)

\bibitem{verdepol05}
Verde, L., Peiris, H.~V. \& Jimenez, R.\ 2006, Journal of Cosmology and Astro-Particle Physics, 1, 19 


\bibitem{Fisher}
R.A.~Fisher, 
\newblock J. Roy. Stat. Soc., {\bf 98}, 39 (1935)

\bibitem{Heavens-Kitching-Verde07}
A. F.~Heavens, T.D.~Kitching and L.~Verde
\newblock Mon. Not. Roy. Astron. Soc. {\bf 380},1029-1035 (2007)


\bibitem{cosmoMC}
A.~Lewis  and S.~Bridle,
\newblock Phys. Rev. D {\bf 66}, 103511 (2002)
and
http://cosmologist.info/cosmomc/

\bibitem{Getdist}
http://cosmologist.info/cosmomc/readme.html\#Analysing



\bibitem{zd-camb-Hamann}
J.~Hamann, S.~Hannestad, J.~Lesgourgues, C.~Rampf and Y. Y.~Wong
\newblock astro-ph:1003.3999v1


\bibitem{Boss}
BOSS http://www.sdss3.org

\bibitem{Euclid}
EUCLID http://sci.esa.int/euclid 

\bibitem{LRMVJ} 
Lopez Honorez, L., Reid, B.~A., Mena, O., Verde, L., \& Jimenez, R.\ 2010, arXiv:1006.0877 


\bibitem{AcquavivaVerde07}
Acquaviva, V., \& Verde, L.\ 2007, Journal of Cosmology and Astro-Particle Physics, 12, 1 


\bibitem{Ishak06}
 Ishak, M., Upadhye, A., \& Spergel, D.~N.
 \ 2006, \prd, 74, 043513 



\bibitem{Hu-Sug-Silk95}
W. ~Hu, N. ~Sugiyama, J. ~Silk,
\newblock Nature {\bf 386}, 37-43 (1997)

\end{thebibliography}
\end{document}